\documentclass[aip,twocolumn,letterpaper]{revtex4}

\usepackage{epsfig}
\usepackage{xcolor}
\definecolor{ultramarine}{rgb}{0.07, 0.04, 0.56}
\usepackage{graphics}
\usepackage{epsfig}

\usepackage[margin=1.0in]{geometry}

\begin{document}
\title{Pfirsch-Schl\"uter Current } 
\author{Allen H Boozer}
\affiliation{Columbia University, New York, NY  10027 \linebreak ahb17@columbia.edu}

\begin{abstract} 
The Pfirsch-Schl\"uter current is a current that flows along the magnetic field lines in a toroidal plasma equilibrium that is required to make the plasma current density divergence free in the presence of a plasma-pressure gradient.  A  distortion in the plasma shape is caused by the Pfirsch-Schl\"uter current, and it is desirable to minimize both the strength and the distance this current flows along the magnetic field lines.   The Pfirsch-Schl\"uter current is localized within a half period of a stellarator when $d\ell/B$ integrated over the half period is the same for all lines in the magnetic surface.  It is shown that within parts in a thousand this is the same condition as the distance $\ell_{s}$ required for a field line to cross the half period being the same for all lines in the surface.  To make the $\ell_{s}$'s the same, the lines started on the small major radius side of the plasma must undergo wiggles to make their $\ell_{s}$ as long as those started on the outboard side.  This is generally achieved using modular coils with a large helical component on the small major radius side but could be achieved with a central column carrying a helical current.

 \end{abstract}

\date{\today} 
\maketitle

\section{Introduction} 

In 1962, Pfirsch and Schl\"uter  \cite{P-S} derived an expression for the current that flows along the magnetic field lines in a toroidal plasma equilibrium that is required to make the plasma current density divergence free in the presence of a plasma-pressure gradient.

A general expression for the Pfirsch-Schl\"uter current is derived in Section V.A of \cite{Boozer:RMP}.   In axisymmetric equilibria, the Pfirsch-Schl\"uter current is inversely proportional to the rotational transform, $\iota$, which is itself proportional to the poloidal divided by the toroidal magnetic field strength.  Consequently, systems,  such as the power-plant design that is based on the levitated toroidal dipole  \cite{OpenStar}, that have only a poloidal field have no Pfirsch-Schl\"uter current.  Even a toroidal dipole requires a non-zero toroidal field in order to avoid problems caused by tiny toroidal asymmetries \cite{Toroidal Dipole}, but the Pfirsch-Schl\"uter current remains sufficiently small to be of no relevance. 

As explained in Section V.A of \cite{Boozer:RMP} and illustrated in \cite{Eq-beta-lim}, the Pfirsch-Schl\"uter current causes a distortion in the plasma shape, which limits the plasma pressure $p$, or more precisely the $\beta\equiv 2\mu_0 p/B^2$, for which a plasma equilibrium is possible.   The Pfirsch-Schl\"uter current also complicates the optimization of stellarators that is required to obtain acceptable neoclassical transport.  To minimize these effects, it is desirable to minimize both the strength and the distance that the Pfirsch-Schl\"uter current flows along the magnetic field lines \cite{Schluter:1983}.  

The degree to which the Pfirsch-Schl\"uter current can be suppressed is not clear while retaining other desirable properties of a stellarator \cite{Wobig:1999}.  This issue of suppression in stellarators was reviewed by Sato et al \cite{PS supp}, and they presented and example of a stellarator with zero Pfirsch-Schl\"uter current. 

One would like to limit the distance that Pfirsch-Schl\"uter currents flow along the magnetic field lines to a segment in the toroidal angle, $\varphi_s\equiv 2\pi/N_s$, where  $N_s$ is an integer.  This integer is equal to either the number of periods $N_p$ or twice the number of periods of the stellarator.   This reduction is more applicable to quasi-isodynamic stellarators, which were discussed by Schl\"uter \cite{Schluter:1983} and Wobig \cite{Wobig:1999}, than to quasi-symmetric stellarator in which the form of Pfirsch-Schl\"uter current \cite{Boozer:1981} implies a scaling as $1/\iota$ in quasi-axisymmetry or as $1/(N_p-\iota)$ in quasi-helical symmetry.

The magnetic field can always be written in Clebsch coordinates $(\psi_t,\theta_0,\ell)$ in which $\vec{B} = \vec{\nabla}\psi_t\times\theta_0/2\pi$ and $d\ell$ is the differential distance along the magnetic field line.  Each magnetic field line is designated by a specific value of $\psi_t$, the magnetic flux enclosed by a magnetic surface, and the Clebsch angle $\theta_0$.  Coordinate systems will be discussed in Section \ref{Sec:C-S}.

The focus of this paper is not on the degree to which the effects of the Pfirsch-Schl\"uter current can be reduced but on showing that the distance $\ell_s(\psi_t,\theta_0)$ that a field line requires to cross the toroidal segment is a surprisingly important parameter in determining whether the Pfirsch-Schl\"uter current remains in the segment. 

\color{black} 
It will be assumed that the net toroidal current in the plasma $I$ is small compared to the poloidal current $G$ in the coils that produces the toroidal magnetic flux.  As shown in Section \ref{sec:P-S}, in the the limit in which $|I/G|<<1$ the \color{black} condition is that $v_s(\psi_t,\theta_0) \equiv \int_0^{\ell_s} d\ell/B$ be independent of the Clebsch angle $\theta_0$.  The surprising result is that the independence of $v_s$ from the Clebsch angle is almost equivalent to the independence of $\ell_s$ from the Clebsch angle.

As will be shown in Section \ref{QI}, the reason that the constancy of $\ell_s$ in $\theta_0$ is almost equivalent to the constancy of $v_s$ comes from the relation between the current and the magnetic field,
\begin{eqnarray}
 && \int_0^{\ell_s} B d\ell= \mathcal{I}_s(\psi_t), \mbox{    where   } \label{B-dl} \\
 && \mathcal{I}_s \equiv \mu_0\frac{G+\iota I}{N_s}. \label{math-I}
 \end{eqnarray}
 The poloidal current outside a magnetic surface is $G(\psi_t)$, the toroidal current enclosed by a magnetic surface is $I(\psi_t)$, and $\iota(\psi_t)$ is the rotational transform.   The dimensionless quantity 
\begin{equation}
\sigma_s \equiv \frac{\Big(\int_0^{\ell_s} \frac{d\ell}{B}\Big)\Big( \int_0^{\ell_s} B d\ell \Big)}{\ell_s^2} \label{sig_s}
\end{equation}
is close to unity and even $\sigma_s-1$ has only a small fractional variation with $\theta_0$.   The exact relation $v_s = \sigma_s \ell_s^2/\mathcal{I}_s$ implies that when $\sigma_s$ has only a weak dependence on $\theta_0$, the variation of $\ell_s^2$ with $\theta_0$ is essentially equivalent to the variation of $v_s$.   

Section \ref{QI} shows that the dependence of $v_s$ on $\theta_0$ is given by the dependence of $\ell_s$ remains correct to parts in a thousand even in quasi-isodynamic stellarators in which the magnetic field strength undergoes an approximate factor of 1.7 change in strength along a field line within a single toroidal segment.
  
\begin{figure}
\centerline{ \includegraphics[width=2 in]{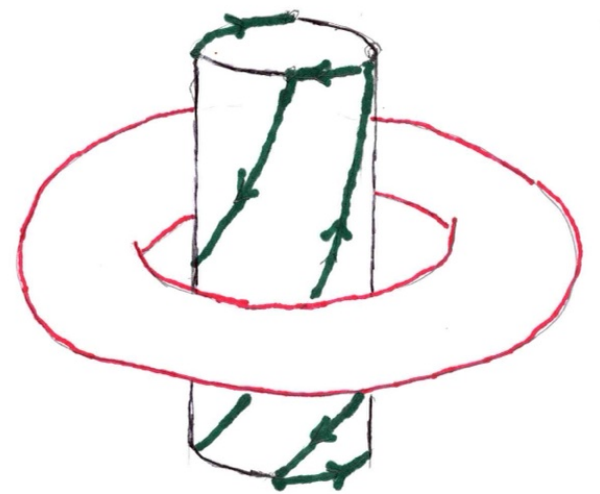}}
\caption{ The coils for optimized stellarators have a strong helical component on the small major radius side.  This is in part to lengthen the magnetic field lines on that side.  The usual solution is to use modular coils that have a strong helical shaping on the small major radius side.  The magnetic field produced by this helical shaping could be produced by a helical winding on a central column.  This should be simpler to design than the central solenoid of tokamaks due to the larger aspect ratio of stellarators. } 
\label{fig:helical-post}
\end{figure}

To make the $\ell_s$'s independent of $\theta_0$, the lines started on the small major radius side of the plasma must undergo wiggles to make their $\ell_s$ as long as those started on the outboard side.  This requires a strong helical field on the small major radius side of a stellarator, which is generally achieved using modular coils \cite{Modulars:1972} but could be achieved by a helical winding on a post, Figure \ref{fig:helical-post}.  

A helical post is similar in concept to the toroidally directed coils of the central solenoid of tokamaks.  Furth and Hartman \cite{Furth-Hartman} have published a related concept for stellarator coils.  The concept was the imposition a strong helical field on the large major radius side, which increases the depth of the stabilizing magnetic well.  Their 1968 publication was long before the optimization of stellarators to minimize the Pfirsch-Schl\"uter current and the neoclassical transport, which require the external helical field be stronger on the small major radius side.


Section \ref{Discussion} summarizes and discusses the results.

In the late 1980's, Arnulf Schl\"uter pointed out to me the approximate condition for the avoidance of a large scale Pfirsch-Schl\"uter currents---all the field lines that lie on a magnetic surface must have the same length as they traverse a period.   As was typical of Schl\"uter, he made the point verbally without an indication of a written reference.  In a 1983 paper \cite{Schluter:1983},  Schl\"uter had pointed out the importance of $\int d\ell/B$ for the maximization of the plasma pressure that can be confined by a stellarator.  A 1973 paper on plasma equilbria in which each magnetic field line closes on itself by Todoroki \cite{Todoroki:1976} considered the condition on both $\oint d\ell/B$ and $\oint B d\ell$ as well as providing an historical review of the subject.   My interest was rekindled by Rohan Ramasamy of Proxima Fusion, who wanted to better understand what limits the reduction of the effects of Pfirsch-Schl\"uter currents.


\section{Coordinate Systems \label{Sec:C-S} }

The theory of general coordinate systems is used in this paper.  The manipulations are simple, though not generally taught.  They are all derived in a two-page appendix to Reference \cite{Boozer:RMP}.

In equilibrium plasmas in which the constant-pressure surfaces are toroidal, the magnetic field can be written as \cite{Boozer:1981}
\begin{eqnarray}
\vec{B} &=&\vec{\nabla}\psi_t\times\vec{\nabla}\frac{\theta}{2\pi} + \iota(\psi_t) \vec{\nabla}\frac{\varphi}{2\pi} \times \vec{\nabla}\psi_t \label{contra}\\
&=& \mu_0 G(\psi_t) \vec{\nabla}\frac{\varphi}{2\pi} + \mu_0I(\psi) \vec{\nabla}\frac{\theta}{2\pi} + \beta_*\vec{\nabla}\psi_t. \hspace{0.2in}\label{co}.
\end{eqnarray}
The pressure is given as $p(\psi_t)$, where $\psi_t$ is the toroidal magnetic flux enclosed by a constant-pressure surface, $\theta$ and $\varphi$ are special choices for the poloidal and the toroidal angles, and $\iota(\psi_t)$ is the twist or rotational transform of the magnetic field lines around the torus.

The discussion of Pfirsch-Schl\"uter currents will use two closely relates sets of Clebsch coordinates, $(\psi_t,\theta_0,\varphi)$ and $(\psi_t,\theta_0,\ell)$  Both $\psi_t$ and  $\theta_0$ are constant along a magnetic field line and provide a label for each line.   The differential distance along each magnetic field line is $d\ell$,
\begin{equation}
\vec{B}\cdot\frac{\partial \vec{x}(\psi_t,\theta_0,\ell)}{\partial\ell} = B.  \label{ell-distance}
\end{equation}
The Clebsch angle and the representation of Equation (\ref{contra}) for $\vec{B}$ in both coordinate sets are
\begin{eqnarray}
\theta_0 &\equiv&\theta - \iota \varphi  \hspace{0.2in} \mbox{   and   }\\
\vec{B}&=&\vec{\nabla}\psi_t\times\vec{\nabla}\frac{\theta_0}{2\pi} \label{Contra}.
\end{eqnarray}

The representation of Equation (\ref{co}) for $\vec{B}$ is slightly more complicated to write even in $(\psi_t,\theta_0,\varphi)$ coordinates.  
\begin{eqnarray}
\vec{B}&=&\frac{\mu_0G}{2\pi} \vec{\nabla}\varphi + \frac{\mu_0I}{2\pi} \vec{\nabla}\theta(\psi_t,\theta_0,\varphi) + \beta_*\vec{\nabla}\psi_t;\\
&=& \frac{\mu_0 (G+\iota I)}{2\pi} \vec{\nabla}\varphi + \frac{\mu_0I}{2\pi} \vec{\nabla}\theta_0 + B_\psi \vec{\nabla}\psi_t. \label{Co} 
\end{eqnarray}
The relation between $\varphi$ and $\ell$ is obtained by dotting Equation (\ref{Co}) with $(\partial \vec{x}/\partial\ell)_{\psi_t \theta_0}$ and using Equation (\ref{ell-distance}), 
\begin{eqnarray}
\Big(\frac{\partial \varphi}{\partial \ell}\Big)_{\psi_t \theta_0} &=& \frac{2\pi B}{\mu_0 (G+\iota I)}. \label{varphi-ell}
\end{eqnarray}
Equation (\ref{varphi-ell})  gives the expression $\int_0^{\ell_s} Bd\ell=\mathcal{I}_s$, Equation (\ref{B-dl}), with $\mathcal{I}_s\equiv  \mu_0(G+\iota I)/N_s$, Equation (\ref{math-I}).

The fact that $d\ell$ is the differential distance along a magnetic field line may be obvious.  A proof follows from the theory of general coordinates
\begin{eqnarray}
\frac{\partial \vec{x}}{\partial\ell} &=& \frac{\vec{\nabla}\psi_t\times\vec{\nabla}\theta_0}{(\vec{\nabla}\psi_t\times\vec{\nabla}\theta_0)\cdot\vec{\nabla}\ell }\\
&=& \frac{\vec{B}}{B}
\end{eqnarray}
using Equations (\ref{ell-distance}) and (\ref{Contra}).


\section{Pfirsch-Schl\"uter  current and $\int \frac{d\ell}{B}$ \label{sec:P-S} }

The equilibrium equation $\vec{\nabla}p=\vec{j}\times \vec{B}$ gives the current density perpendicular to the magnetic field, $\vec{j}_\bot$, which is generally not divergence free.  The condition that $\vec{\nabla}\cdot\vec{j}=0$ implies 
\begin{eqnarray}
\vec{\nabla}\cdot\vec{j} &=& \vec{\nabla}\cdot\Big(\frac{j_{||}}{B}\vec{B}\Big) + \vec{\nabla}\cdot\vec{j}_\bot=0.\\
\vec{j}_\bot &=& \frac{dp(\psi_t)}{d\psi_t} \frac{\vec{B}\times\vec{\nabla}\psi_t}{B^2}  \\
&=& \Big(\frac{dp}{d\psi} \frac{\vec{B}}{B^2}\Big)\times \vec{\nabla}\psi_t. \\
\vec{\nabla}\cdot(\vec{A}\times\vec{B}) &=& \vec{B}\cdot \vec{\nabla}\times \vec{A}  - \vec{A}\cdot \vec{\nabla}\times \vec{B},  \mbox{   so   } \\
\vec{\nabla}\cdot\vec{j}_\bot &=& \vec{\nabla}\psi_t \cdot \vec{\nabla}\times\Big(\frac{dp}{d\psi_t} \frac{\vec{B}}{B^2}\Big). 
\end{eqnarray}
Since $\vec{\nabla}\psi_t\cdot\vec{\nabla}\times\vec{B}=0$,
\begin{eqnarray}
\vec{\nabla}\cdot\vec{j}_\bot &=& \frac{dp}{d\psi_t}  \vec{\nabla}\psi_t \cdot \Big(\vec{\nabla}\frac{1}{B^2}\times\vec{B}\Big) \\
&=& \frac{dp}{d\psi_t} \Big(\vec{B} \times \vec{\nabla}\psi_t\Big) \cdot \vec{\nabla}\frac{1}{B^2}.  \mbox{   but   } \\
\vec{B} \times \vec{\nabla}\psi_t &=& \frac{1}{2\pi} \Big(\mu_0(G+\iota I) \vec{\nabla}\varphi\times\vec{\nabla}\psi_t \nonumber\\&& -\mu_0I \vec{\nabla}\psi_t\times\vec{\nabla}\theta_0\Big)\\
&=&(\vec{B}\cdot\vec{\nabla}\varphi)  \Big(\mu_0(G+\iota I) \frac{\partial\vec{x}}{\partial\theta_0} \nonumber\\&& -\mu_0I \frac{\partial\vec{x}}{\partial\varphi}\Big) \label{B-grad-psi}\\
\frac{\vec{\nabla}\cdot\vec{j}_\bot}{\vec{B}\cdot\vec{\nabla}\varphi} &=& \frac{dp}{d\psi_t} \Big(\mu_0(G+\iota I)\frac{\partial(1/B^2)}{\partial\theta_0} \nonumber\\&& - \mu_0I \frac{\partial(1/B^2)}{\partial\varphi} \Big), \mbox{    but   } \\
\vec{\nabla}\cdot\Big(\frac{j_{||}}{B}\vec{B}\Big) &=& \vec{B}\cdot\vec{\nabla}\varphi \frac{\partial}{\partial\varphi} \frac{j_{||}}{B} \\
&=& - \vec{\nabla}\cdot\vec{j}_\bot,  \mbox{    so   } \\
\frac{\partial}{\partial\varphi} \frac{j_{||}}{B} &=&-\frac{dp}{d\psi_t} \Big(\mu_0(G+\iota I)\frac{\partial(1/B^2)}{\partial\theta_0} \nonumber\\&& + \mu_0I \frac{\partial(1/B^2)}{\partial\varphi} \Big). \hspace{0.2in}  \mbox{   Since   }\\
\frac{\partial}{\partial\varphi}&=& \frac{\mu_0(G+\iota I)}{2\pi B} \frac{\partial}{\partial\ell},\\
\frac{\partial}{\partial\ell} \frac{j_{||}}{B} &=&-2\pi B\frac{dp}{d\psi_t} \frac{\partial(1/B^2)}{\partial\theta_0} \nonumber\\&& + \frac{dp}{d\psi_t} \mu_0I \frac{\partial(1/B^2)}{\partial\ell} \\
&=&-4\pi \frac{dp}{d\psi_t} \frac{\partial(1/B)}{\partial\theta_0} \nonumber\\&& + \frac{dp}{d\psi_t} \mu_0I \frac{\partial(1/B^2)}{\partial\ell} \label{partial j_||/B} \label{j_||/B Eq}
\end{eqnarray}

The density of parallel current $\vec{j}_{||} = (j_{||}/B)\vec{B}$ is the sum of two parts: the Pfirsch-Schl\"uter current density $\vec{j}_{PS} = (j_{PS}/B)\vec{B}$, which has a divergence that balances the divergence of the confining current density $\vec{j}_\bot$ and a divergence-free net current density $(j_{net}/B)\vec{B}$. \color{black} Equation (64) in \cite{Boozer:RMP}  gives
\begin{eqnarray}
\frac{j_{net}}{B} &=& \frac{G \frac{dI}{d\psi_t} - I \frac{dG}{d\psi_t} }{G+\iota I} \label{j-net}\\
&\approx \frac{dI}{d\psi_t} 
\end{eqnarray}
when $|I/G|<<1$.  


Equation (\ref{partial j_||/B}) can be written as 
\begin{eqnarray}
&&\frac{\partial}{\partial\ell} \Big(\frac{j_{||}}{B}- \frac{dp}{d\psi_t} \frac{\mu_0I}{B^2}  + h(\psi)\Big) \nonumber\\ && \hspace{1.0in}= -4\pi \frac{dp}{d\psi_t} \frac{\partial(1/B)}{\partial\theta_0}. \hspace{0.3in} \label{h-eq}
\end{eqnarray}
The homogeneous solution, $h(\psi)$, to this partial differential equation is needed to ensure
\begin{eqnarray}
\lim_{L\rightarrow\infty} \frac{1}{L}\int_0^L \frac{j_{||}}{B}d\ell &=& \frac{G \frac{dI}{d\psi_t} - I \frac{dg}{d\psi_t}}{G+\iota I} \\ 
&\approx&  \frac{dI}{d\psi}, \mbox{ when  } \left|\frac{I}{G}\right| <<1. \hspace{0.2in} \label{j_net}
\end{eqnarray}
The remainder of $j_{||}/B$ gives the Pfirsch-Schl\"uter current density $(j_{PS}/B)\vec{B}$.  The fundamental equation for the Pfirsch-Schl\"uter current is then
\begin{eqnarray}
\frac{\partial}{\partial\ell} \frac{j_{PS}}{B} &=&-4\pi \frac{dp}{d\psi_t} \frac{\partial(1/B)}{\partial\theta_0},  \label{P-S}
\end{eqnarray}
when $|I/G|<<1$.


Wobig's \cite{Wobig:1999} gave an expression for the Pfirsch-Schl\"uter current, $(j_{PS}^W/B)\vec{B}$, using the geodesic curvature of the magnetic field lines
\begin{eqnarray}
\kappa_g &=& (\vec{B}\times\hat{n})\cdot\vec{\nabla}\frac{1}{B}. \\
\vec{B}\cdot\vec{\nabla}\frac{j_{PS}^W}{B} &=&-2 \frac{dp}{d\psi_t} \frac{|\vec{\nabla}\psi_t|}{B}\kappa_g,
\end{eqnarray}
where $\hat{n}=\vec{\nabla}\psi_t/ |\vec{\nabla}\psi_t|$.  Using Wobig's expressions and Equation (\ref{B-grad-psi}),
\begin{eqnarray}
\frac{\partial}{\partial\ell} \frac{j_{PS}^W}{B} &=& -2\frac{dp}{d\psi_t} \frac{\vec{B}\times\vec{\nabla}\psi_t}{B^2} \cdot\vec{\nabla}\frac{1}{B},
\end{eqnarray} 
which gives the equation
\begin{eqnarray}
\frac{\partial}{\partial\ell} \frac{j_{PS}^W}{B} &=&-4\pi \frac{dp}{d\psi_t} \frac{\partial(1/B)}{\partial\theta_0} \nonumber\\&& + \frac{dp}{d\psi_t} \mu_0I \frac{\partial(1/B^2)}{\partial\ell}, 
\end{eqnarray}
 which is identical form to Equation (\ref{j_||/B Eq}) when $j_{||}/B$ is replaced by $j_{PS}^W/B$.  \color{black}

 The net plasma current $I(\psi_t)$ is zero in a stellarator that has no bootstrap current or other current drive mechanisms, but the Pfirsch-Schl\"uter current is not.

\color{black}

Equation \ref{j_||/B Eq} implies that the exact change in $j_{||}{B}$  from the beginning to the end of a toroidal segment of length $\ell_s$, which is equivalent to the change in the Pfirsch-Schl\"uter current, is
\begin{eqnarray}
\Delta\left(\frac{j_{||}}{B}\right) &=& \frac{dp}{d\psi_t}\mu_0I \Delta\left( \frac{1}{B^2}\right) -4\pi \frac{dp}{d\psi_t} \frac{\partial v_s}{\partial \theta_0}; \label{full-j_ps} \hspace{0.2in} \\
\Delta\left(\frac{1}{B^2}\right) &\equiv& \frac{1}{B^2(\psi_t,\theta_0,\ell_s)} - \frac{1}{B^2(\psi_t,\theta_0,0)}; \\
v_s(\psi_t,\theta_0) &\equiv& \int_0^{\ell_s} \frac{d\ell}{B}. \label{Eq:v_s}
\end{eqnarray}

The magnitude of first term on the right hand side of Equation \ref{full-j_ps} divided by the second is very small compared to unity
\begin{eqnarray}
\left|\frac{\mu_0I \Delta\left( \frac{1}{B^2}\right)}{4\pi \frac{\partial v_s}{\partial \theta_0}} \right| &\approx& \left|\frac{I}{G} \frac{R_0 \Delta B}{ B_0^2  \int_0^{\ell_s} \frac{d\ell}{B}}\right|,
\end{eqnarray}
where a typical major radius and magnetic field strength are related by $B_0=\mu_0G/2\pi R_0$ and $(\Delta B)/B_0$ is the fractional variation in the field strength from field line to field line on $(\psi_t,\theta_0)$ surfaces separated by the distance $\ell_s$, which as discussed in Section \ref{QI} is not large even when the magnetic field strength has a large variation within a half period.

Equation \ref{full-j_ps} for the exact variation in the Pfirsch-Schl\"uter current is accurately localized within each toroidal segment, $\varphi_s=2\pi/N_s$ of length $\ell_s$ along the field lines when
\begin{eqnarray}
v_s(\psi_t,\theta_0) \equiv \int_0^{\ell_s} \frac{d\ell}{B} \label{Eq:v_s}
\end{eqnarray}
is independent of $\theta_0$, which is the label for the different magnetic field lines that lie in a given constant-pressure surface, which is also a magnetic surface that encloses toroidal flux $\psi_t$.  \color{black}


As noted by Schl\"uter \cite{Schluter:1983}, for an equilibrium to exist the integral $\oint d\ell/B$ must be zero for every closed field line, which means for every field line on surfaces on which $\iota$ is rational number.   Boozer \cite{Boozer:1981} showed that this condition is only of practical relevance on low-order rational surfaces.  Boozer \cite{Boozer:1981} showed that this condition is only of practical relevance on low-order rational surfaces.

Differential equations of the form of Equation (\ref{h-eq}) are called ``magnetic differential equations" by Kruskal and Kulsrud \cite{Kruskal-Kulsrud} and appear commonly in the theory of toroidal plasmas. Their example, $\vec{B}\cdot\vec{\nabla}r =s$ has the solution $r = \int (s/B)d\ell$.  The $1/B$ weighting factor essentially gives a volume average within an infinitesimal tube defined by  magnetic field lines on its surface that contains a flux $d\psi_t d\theta_0/2\pi$.  The theory of general coordinates implies the volume  of a surface defined by magnetic field lines is
\begin{eqnarray}
V(\psi_t) &=& \int \frac{d\psi_t d\theta_0 d\ell}{(\vec{\nabla}\psi_t\times\vec{\nabla}\theta_0)\cdot\vec{\nabla}\ell }\\
&=& \int \frac{d\psi_t d\theta_0 d\ell}{2\pi B }\\
&=& \frac{1}{2\pi}\int \Big(\frac{ d\ell}{B }\Big)d\psi_t d\theta_0 \\
&=& \frac{N_s}{2\pi} \int v_s(\psi_t,\theta_0) d\psi_t d\theta_0,
\end{eqnarray}
where $N_s$ is the number of toroidal segments.  This means that $v_s(\psi_t,\theta_0)$ is the volume within a toroidal segment of an infinitesimal tube defined by magnetic field limes that contains a magnetic  flux  $d\psi d\theta_0/2\pi$.  The magnetic flux in a tube defined by field lines is 
\begin{eqnarray}
\mbox{Flux} &=& \int \frac{\vec{B}\cdot\vec{\nabla}\ell}{(\vec{\nabla}\psi_t\times\vec{\nabla}\theta_0)\cdot\vec{\nabla}\ell }d\psi_t d\theta_0\\
&=& \frac{1}{2\pi}\int d\psi_t d\theta_0 =  \frac{d\psi_t d\theta_0 }{2\pi}
\end{eqnarray}
for an infinitesimal tube.


As will be shown in Section \ref{QI}, the condition that all magnetic field lines in a toroidal segment $\varphi_s$ must have the same length to keep the Pfirsch-Schl\"uter currents confined to the segment is an extremely accurate approximation even in the precise quasi-isodynamic stellarator of Goodman et al \cite{Goodman}  in which the maximum magnetic field strength along a line within a half-period is 1.7 times the minimum.


\section{Sensitivity of the $\ell_s$ condition to the non-constancy of $B$ \label{QI} }

No matter how large the variation in the magnetic strength along a field line may be,  Equations (\ref{Eq:v_s}) and (\ref{math-I}) and the definition of $\sigma_s$, Equation (\ref{sig_s}), can be used to write
\begin{eqnarray}
v_s(\psi_t,\theta_0) &=&\sigma_s(\psi_t,\theta_0) \frac{\ell_s^2}{\mathcal{I}_s} \mbox{    or  } \label{v_s}\\
\sigma_s(\psi_t,\theta_0)&\equiv& \frac{v_s \mathcal{I}_s}{\ell_s^2}.  
\end{eqnarray}
When $\sigma_s$ is approximately unity and has only a week dependence on $\theta_0$, the condition that all magnetic field lines in a segment must have the same length to keep the Pfirsch-Schl\"uter currents confined to a segment is an extremely accurate approximation. 


\subsection{Maximum deviation from of $\sigma_s$ from unity \label{Sec:max-dev}}

The maximum possible deviation of $\sigma_s(\psi_t,\theta_0)$ from unity is given by assuming the field strength has its largest value, $B_\ell$, for a distance $\ell_\ell$ and its smallest value, $B_{sm}$, for a distance $\ell_s-\ell_\ell$.  Then,
\begin{eqnarray}
v_s &=& \frac{\ell_\ell}{B_\ell} + \frac{\ell_s - \ell_\ell}{B_{sm}} \\
&=& \frac{\ell_s}{B_{sm}} \Big(1 - \frac{\ell_\ell}{\ell_s} \frac{B_\ell -B_{sm}}{B_\ell} \Big). \\ 
\mathcal{I}_s &=& B_\ell \ell_\ell + B_{sm} (\ell_s-\ell_\ell) \\
&=& B_{sm} \ell_s \Big(1 + \frac{\ell_\ell}{\ell_s}\frac{B_\ell}{B_{sm}}  \frac{B_\ell -B_{sm}}{B_\ell} \Big) \mbox{   so   } \\
\frac{1}{B_{sm}} &=& \frac{\ell_p}{\mathcal{I}} \Big(1 + \frac{\ell_\ell}{\ell_s}\frac{B_\ell}{B_{sm}}  \frac{B_\ell -B_{sm}}{B_\ell} \Big).
\end{eqnarray}

Since $v_s=\sigma_s \ell_s^2/\mathcal{I}$, an inequality holds: $\sigma_s\leq \sigma_s^{max}$ with 
\begin{eqnarray}
\sigma_s^{max} &\equiv&1+\Big\{\frac{\ell_\ell}{\ell_s} \Big(1 - \frac{\ell_\ell}{\ell_s}\Big)\Big\} \frac{(B_\ell - B_{sm})^2}{B_\ell B_{sm}}. \label{sigma_s-max}
\end{eqnarray} 
The actual $\sigma_s(\psi_t,\theta_0)$ must be closer to unity than is $\sigma_s^{max}$.  In addition, the factor $\sigma_s^{max}-1$ varies little from field line to field line as it crosses either a half or a whole period.  When $\sigma_s(\psi_t,\theta_0)$ is almost independent $\theta_0$, the constancy of $\ell_s$ essentially ensures the Pfirsch-Schl\"uter current is confined to each half or whole period.

Optimized designs for quasi-isodynamic stellarators have much larger values of the mirror ratio $(B_\ell - B_{sm})^2)/(B_\ell B_{sm})$ than for quasi-symmetric stellarators.  Quasi-isodynamic stellarators can have better confinement t \color{black} with a \color{black} larger mirror ratio and are sometimes called linked mirrors.  When $\sigma_s(\psi_t,\theta_0)$ is almost independent of $\theta_0$ for both quasi-isodynamic and quasi-symmetric stellarators, the variation in $\ell_s$ between lines is a good measure of the variation in $v_s$ with $\theta_0$.

As an example, consider the precise quasi-isodynamic stellarator of Goodman et al \cite{Goodman}.  Figure 3 of that paper showed the magnetic field strength was approximately sinusoidal within a period and varied between approximately 3.9~T and 6.6~T.   Take  $B_\ell = 6.6~$T and $B_{sm}=3.9~$T, so $(B_\ell - B_{sm})^2/(B_\ell B_{sm}) =0.283$.  When $\ell_\ell=\ell_s/2$, $\sigma_s = 1.0708$.  Figure 2 of Goodman et al \cite{Goodman} shows that the well shape has a full range of shape changes of $\approx 20\%$ from field line to field line, so $\sigma \approx 1.0708\pm 0.0035$. If $v_s$ were independent of $\theta_0$, one finds  $\ell_s^2 \approx (1.0708 \pm 0.007)$, which would imply $\ell_s$ is constant to within four parts in a thousand.


\subsection{More accurate full-period result}

The magnetic field stength $B$ in the quasi-isodynamic stellarator of Goodman et al \cite{Goodman} is $B\approx B_0(1+a_m \sin(2\pi \ell/\ell_p))$ with $a_m\approx(6.6 - 3.9)/(6.6+3.9)\approx0.26$.  Equation (\ref{sig_s}) applied over a full period gives $\sigma_p$.  The two integrals are
\begin{eqnarray}
&&\int_0^{2\pi} \frac{d\alpha}{1+a_m \sin\alpha} =\frac{2\pi}{\sqrt{1-a_m^2} } \\
&& \int_0^{2\pi} (1+a_m \sin\alpha)d\alpha = 2\pi \\
&&\sigma_p \approx  \frac{1}{\sqrt{1-a_m^2} } \\
&& \hspace{0.15in} \approx 1.036,
\end{eqnarray}
which is approximately half the deviation from unity as the maximum deviation obtained in Section \ref{Sec:max-dev}.  Figure 2 of \cite{Goodman} shows that the well shape changes by approximately 20\% from field line to field line, so $\sigma \approx 1.036\pm 0.004$.  The implication is that the condition for the Pfirsch-Schl\"uter to remain within a period is \color{black} that the length of all field lines in a period $\ell_p$ be a constant, which  \color{black} is correct to within two parts in a thousand. \\


\subsection{More accurate half-period result}

Equation (\ref{sig_s}) applied over a half period gives $\sigma_{p/2}$.  The two integrals are

\begin{eqnarray}
&&\int_0^{\pi} \frac{d\alpha}{1+a_m \sin\alpha} =\frac{\pi \Big(1-\frac{2 \arcsin a_m}{\pi}\Big)}{\sqrt{1-a_m^2} } \\
&& \int_0^{2\pi} (1+a_m \sin\alpha)d\alpha = \pi + 2 a_m\\
&&\Big(\int_0^{\pi} \frac{d\alpha}{1+a_m \sin\alpha}\Big)\Big( \int_0^{2\pi} (1+a_m \sin\alpha)d\alpha\Big)\nonumber \\&& \hspace{0.5in}=\frac{0.97036\pi^2}{\sqrt{1-a_m^2} }\\
&&\sigma_{p/2} \approx \frac{ \Big(1-\frac{2 \arcsin a_m}{\pi}\Big)}{\sqrt{1-a_m^2} }\Big(1+\frac{2a_m}{\pi}\Big)  \hspace{0.2in}\\
&& \hspace{0.15in} \approx 1.0049
\end{eqnarray}

\section{Discussion \label{Discussion}}

Stellarator equilibria naturally have a Pfirsch-Schl\"uter current along the magnetic field lines.  This current has a divergence that cancels the divergence in the current density perpendicular to the magnetic field, $\vec{j}_\bot$, that is required to balance the pressure gradient, $\vec{j}_\bot\times\vec{B}=\vec{\nabla}p$.   The effect of the Pfirsch-Schl\"uter current is to distort the magnetic surfaces from the shape they would have in the externally produced magnetic field.  This complicates the optimization of stellarators and more importantly limits the obtainable $\beta$ in stellarator equilibria \cite{Schluter:1983,Eq-beta-lim}.  

The Pfirsch-Schl\"uter current is defined using the homogeneous solution $h(\psi_t)$ to Equation (\ref{h-eq}) to ensure that the average of $j_{||}/B$ along magnetic field lines \color{black} gives the net plasma current, Equation (\ref{j-net}).  When $|I/G|<<1$, Equation (\ref{P-S}) gives the Pfirsch-Schl\"uter current $(\partial (j_{PS}/B)/\partial \ell)_{\psi_t \theta_0}=-4\pi(dp/d\psi_t) (\partial(1/B)/\partial\theta_0)_{\psi_t \ell}$.   

In general, the variation in the Pfirsch-Schl\"uter current over a toroidal segment of length $\ell_s$ is given by Equation (\ref{full-j_ps}).   This equation implies that when $|I/G|$ is not negligibly small, that the variation in the magnetic field strength along the magnetic field lines in the presence of a pressure gradient and a net plasma current $I(\psi_t)$ makes a contribution to the the Pfirsch-Schl\"uter current in addition to the contribution from the variation of $v_s\equiv\int_0^{\ell_s} d\ell/B$.  

\color{black}

The size of the distortion produced by the Pfirsch-Schl\"uter current increases not only with its magnitude but also with the distance along the magnetic field lines required to obtain a zero average of that current.  Quasi-isodynamic stellarators could be optimized to make this distance a half of a stellarator period.  This optimum is achieved to remarkable accuracy when each magnetic field line has the same length $\ell_s$ as it crosses a half period and achieved exactly when $v_s\equiv\int_0^{\ell_s}d\ell/B$, Equation (\ref{Eq:v_s}), is the same for each line.  The reason the constancy of $\ell_s$ is an accurate approximation to the constancy $v_s$ is explained by the Equation  (\ref{v_s}), $v_s = \sigma_s \ell_s^2/\mathcal{I}_s$ and  an inequality that $\sigma_s$ must be closer to unity than is $\sigma_s^{max}$, Equation (\ref{sigma_s-max}).  But, even the deviation of $\sigma_s$ from unity tends to have only a weak dependence on which field line is followed through a half period when the length $\ell_s$ is the same for all lines. 

Magnetic field lines must be wiggled on the inboard side of a torus to give them the same length as they pass through a whole or a half a period as they have on the outboard side.  Indeed, much stronger helical contributions to the external coil currents on the inboard side are a characteristic feature of optimized stellarator equilibria.  This can be achieved by many coil design concepts.  Figure \ref{fig:helical-post} illustrates one way to achieve this effect that to the best of my knowledge has not yet been studied, but investigating what coil-design concept is optimal is outside of the scope of this paper.

\section*{Acknowledgements}

This material is based upon work supported by the U.S. Department of Energy, Office of Science under Award Nos. DE-SC0024548, and DE-AC02-
09CH11466.

 \vspace{0.01in}

\section*{Author Declarations}

The author has no conflicts to disclose. \vspace{0.01in}


\section*{Data availability statement}

Data sharing is not applicable to this article as no new data were created or analyzed in this study.


 \end{document}